# Multi-functional microwave photonic radar system for simultaneous distance and velocity measurement and high-resolution microwave imaging

Dingding Liang[a,b], Lizhong Jiang[c,d], and Yang Chen[a,b,*]

[a] *Shanghai Key Laboratory of Multidimensional Information Processing, East China Normal University, Shanghai 200241, China*
[b] *Engineering Center of SHMEC for Space Information and GNSS, East China Normal University, Shanghai 200241, China*
[c] *School of Artificial Intelligence and Automation, Huazhong University of Science and Technology, Wuhan 430074, China*
[d] *Shanghai Radio Equipment Research Institute, Shanghai 201109, China*
[*] ychen@ce.ecnu.edu.cn

**ABSTRACT**

A photonic-assisted multi-functional radar system for simultaneous distance and velocity measurement and high-resolution microwave imaging is proposed and experimentally demonstrated by using a composite transmitted microwave signal of a single-chirped linearly frequency-modulated (LFM) signal and a single-tone microwave signal. In the system, the transmitted signal is generated via photonic frequency up-conversion based on a single integrated dual-polarization dual-parallel Mach-Zehnder modulator (DPol-DPMZM), whereas the echo signals scattered from the target are de-chirped to two low-frequency signals using a microwave photonic frequency mixer. By using the two low-frequency de-chirped signals, the real-time distance and radial velocity of the moving target can be measured accurately according to the round-trip time of the echo signal and its Doppler frequency shift. Compared with the previous reported distance and velocity measurement methods, where two LFM signals with opposite chirps are used, these parameters can be obtained using only a single-chirped LFM signal and a single-tone microwave signal. Meanwhile, high-resolution inverse synthetic aperture radar (ISAR) imaging can also be realized using ISAR imaging algorithms. An experiment is performed to verify the proposed multi-functional microwave photonic radar system. An up-chirped LFM signal from 8.5 to 12.5 GHz and an 8.0 GHz single-tone microwave signal are used as the transmitted signal. The results show that the absolute measurement errors of distance and radial velocity are less than 5.9 cm and 2.8 cm/s, respectively. ISAR imaging results are also demonstrated, which proves the high-resolution and real-time ISAR imaging ability of the proposed system.

## 1. Introduction

Radio detection and ranging (radar) has attracted increasing attention due to its unmatched advantage of operating in all-weather conditions [1]-[2]. To date, radar has been widely used in remote sensing, autonomous vehicles, 3D imaging, and so on [3]-[5]. Different kinds of radar systems can be used to extract various types of information of a target, such as distance, velocity, and target shape [5]-[7]. The distance and velocity information of a target can be obtained by measuring the round-trip time of the scattered signal and its Doppler frequency shift (DFS), while the target shape can be achieved by inverse synthetic aperture radar (ISAR) imaging algorithm [1], [8]. All of this information is of great importance to target detection so that the integration of multiple functions into a signal radar system is of great significance [9]-[11]. To acquire a high-resolution distance and clear target shape, a large-bandwidth and high-frequency radar transmitted signal is highly desired. In conventional radar systems, however, it is hard to generate and process large-bandwidth and high-frequency microwave signals due to the electronic bottlenecks [12], [13], resulting in a low-accuracy measurement of distance and velocity and a blurred target imaging.

During the past few decades, microwave photonics has been extensively investigated due to its unique advantages in generating, transmitting, and processing large bandwidth and high-frequency microwave signals. One of its most important applications is to overcome the bandwidth and frequency limitations of conventional radar systems [14]-[16]. Many photonics-based approaches have been proposed for directly generating and processing high-frequency and large-bandwidth microwave signals [17]-[21], especially the linearly frequency-modulated (LFM) signals [19]-[21]. Based on these technologies, some photonics-based distance and velocity measurement methods have been presented [22]-[24]. Generally, LFM signals with opposite chirps containing up-chirped and down-chirped segments are applied to implement the distance and velocity measurement of a moving target. In [22], a high-resolution range and velocity measurement method was proposed. In the transmitter, a V-shaped LFM microwave signal is generated through the frequency doubling, whereas in the receiver, the information of target distance and velocity can be obtained by analyzing the two beat frequencies generated by the up-chirped and down-chirped segments. In [23], a photonics-based method for simultaneously measuring distance and velocity was proposed using multi-band LFM microwave signals with opposite chirps. In the experiment, a tri-band LFM signal is adopted to determine distance and the magnitude and direction of velocity. The measured relative errors for distance and velocity are less than 0.005% and 0.59%, respectively. To measure the distance and velocity of multiple targets simultaneously, a photonics-based approach was proposed utilizing dual-band symmetrical triangular LFM waveforms [24]. In their experiment, detection of three targets is performed. The absolute measurement errors of the distance and the velocity are less than 9 mm and 0.16 m/s, respectively. These methods all show the advantages of microwave photonics technology in distance and velocity measurement. To meet the increasing demand in military affairs and civilian activities, it is highly urgent to realize the integration of multiple functions into a single radar system [25], [26], such as simultaneously achieving target tracking and imaging. Although the above three methods can achieve distance and velocity measurement [22]-[24], they cannot obtain a clear target shape due to the use of a transmitted signal with opposite chirps. To the best of our knowledge, a multi-functional microwave photonic radar system that can simultaneously achieve distance and velocity measurement and high-resolution ISAR imaging has not yet been described.

In this paper, we propose and demonstrate a multi-functional microwave photonic radar system for distance and velocity measurement and high-resolution microwave imaging. In the radar transmitter, a single integrated dual-polarization dual-parallel Mach-Zehnder modulator (DPol-DPMZM) is used. In the DPol-DPMZM, the +1st-order optical sideband of an intermediate frequency (IF) LFM signal and the optical carrier are generated on the X polarization, while the -1st-order optical sideband of a single-frequency microwave signal is generated on the Y polarization. Then, the X-polarized and Y-polarized light waves are combined and detected in a photodetector (PD), where the -1st-order single-frequency optical sideband is beaten with the carrier and the +1st-order LFM optical sideband to generate a composite transmitted microwave signal of a single-chirped LFM signal and a single-tone microwave signal. In the receiver, the echoes scattered from the target are de-chirped to two low-frequency signals by a microwave photonic frequency mixer. By using the two low-frequency de-chirped signals, the real-time distance and radial velocity of a moving target can be obtained along with high-resolution ISAR imaging. Compared with the previously reported distance and velocity measurement methods using two LFM signals with opposite chirps, the distance and radial velocity can be obtained using only a single-chirped LFM signal and a single-tone microwave signal, which makes it possible for the system to perform high-resolution ISAR imaging in addition to the distance and radial velocity measurement. An experiment is performed. An up-chirped LFM signal from 8.5 to 12.5 GHz and an 8.0 GHz single-tone microwave signal are generated and used as the transmitted radar signal. The results show that the absolute measurement errors of distance and radial velocity are less than 5.9 cm and 2.8 cm/s, respectively. ISAR imaging results are also demonstrated, which proves the high-resolution and real-time ISAR imaging ability of the proposed system.

## 2. Principle

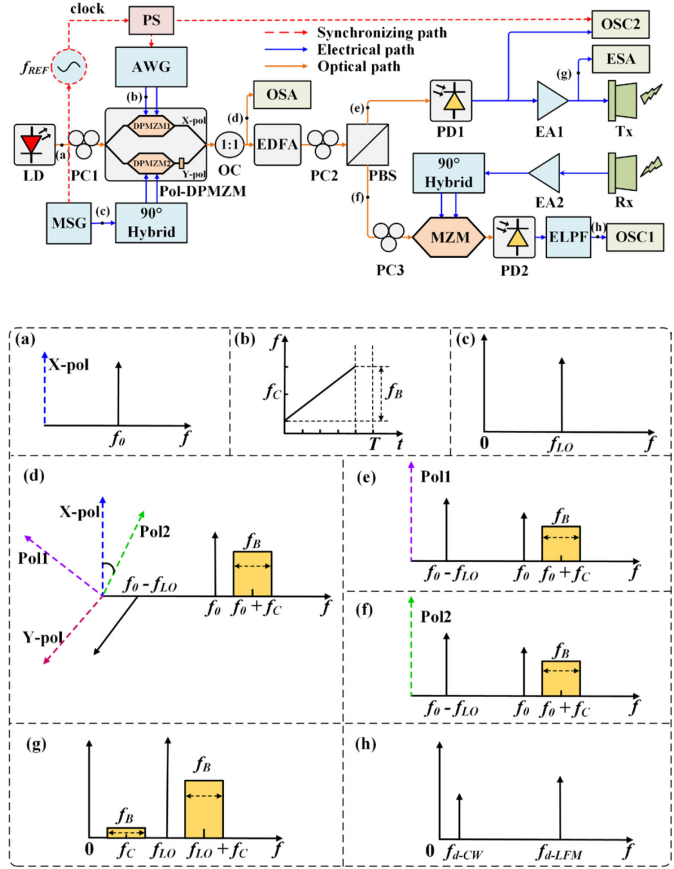

Fig. 1. Schematic diagram of the proposed multi-functional microwave photonic radar system. LD, laser diode; PC, polarization controller; Pol-DPMZM, dual-polarization dual-parallel Mach-Zehnder modulator; OC, optical coupler; EDFA, erbium-doped fiber amplifier; PBS, polarization beam splitter; PD, photodetector; MZM, Mach-Zehnder modulator; OSA, optical spectrum analyzer; AWG, arbitrary waveform generator; MSG, microwave signal generator; PS, power splitter; EA, electrical amplifier; ELPF, electrical low-pass filter; OSC, oscilloscope; Tx: transmitter; Rx: receiver. (a)-(h) are the schematic diagrams of the signals at different locations in the system diagram.

The schematic diagram of the proposed multi-functional microwave photonic radar system is shown in Fig. 1. A continuous-wave (CW) light wave with a center frequency of $f_0$ from a laser diode (LD) is sent to a DPol-DPMZM via a polarization controller (PC1). The DPol-DPMZM consists of two dual-parallel Mach-Zehnder modulators (DPMZMs) in the X and Y polarizations, respectively. DPMZM1 in the X polarization is driven by two phase-orthogonal IF LFM signals with positive chirps, which are generated from a two-channel arbitrary waveform generator (AWG). The instantaneous frequency of the phase-orthogonal LFM signal can be expressed as

$$f_{IF}(t) = \begin{cases} f_c - 0.5f_B + kt, & t \in (0, 0.8T] \\ 0, & t \in (0.8T, T] \end{cases}, \quad (1)$$

where $f_c$ is the center frequency, $f_B$ is the bandwidth, and $k=B/0.8T$ is the chirp rate of the IF LFM signal. The time-frequency diagram of the phase-orthogonal LFM signal is shown in Fig.1(b), where $T$ is the period of the IF LFM signal. The main-MZM and sub-MZMs of DPMZM1 are all biased at the quadrature transmission points to implement the single-sideband (SSB) modulation of the IF LFM signal, as shown in the X polarization in Fig.1(d). A single-tone microwave signal at $f_{LO}$ generated from a microwave signal generator (MSG) is applied to a 90° electrical hybrid coupler and then sent to DPMZM2, which is operated as a carrier-suppressed single-sideband (CS-SSB) modulator. The output of DPMZM2 is shown in the Y polarization in Fig.1(d). Under these circumstances, the output optical signal in the X and Y polarizations of the DPol-DPMZM can be written as

$$E_X(t) \propto J_0(m_1)\exp\left(j2\pi f_0 t + j\frac{\pi}{2}\right) - \sqrt{2}J_1(m_1)\cdot \exp\left[j2\pi\left(f_0 + f_c - \frac{1}{2}f_B\right)t + j\pi kt^2 + j\frac{\pi}{4}\right], \quad (2)$$

$$E_Y(t) \propto J_1(m_2)\exp(j2\pi f_0 t - j2\pi f_{LO}t + j\pi). \quad (3)$$

where $J_n(\cdot)$ is the $n$th-order Bessel function of the first kind and $m_1$ and $m_2$ are the modulation indices of the X-polarized light and Y-polarized light, respectively. Under the small-signal modulation, the higher-order modulation sidebands are ignored in obtaining (2) and (3). At the output port of the DPol-DPMZM, the two orthogonally polarized lights are combined, as shown in Fig.1(d). After the DPol-DPMZM, PC2 and a polarization beam splitter (PBS) are used. By adjusting PC2, the polarization state of the X-polarized light or the Y-polarized light is oriented to have an angle of 45° to one of the principal axes of the PBS. The optical signals from PC2 are split into two branches by the PBS, which are shown in Fig. 1(e) and (f). The two outputs of the PBS can be denoted as

$$E_T(t) \propto E_X(t) + E_Y(t), \quad (4)$$
$$E_R(t) \propto E_X(t) - E_Y(t). \quad (5)$$

In the lower branch, the optical signal $E_R(t)$ is used as a reference signal for de-chirping the radar echo signals, while in the upper branch, the signal $E_T(t)$ is sent to a photodetector (PD1) to perform the optical-electrical conversion.

After detection in PD1, two LFM signals with the same bandwidth and a single-tone microwave signal are generated simultaneously, which can be written as

$$I(t) \propto \sqrt{2}J_1(m_1)J_1(m_2)\cos\left[2\pi\left(f_{LO} + f_c - \frac{1}{2}f_B\right)t + \pi kt^2 + \frac{\pi}{4}\right]$$
$$+ \sqrt{2}J_0(m_1)J_1(m_1)\cos\left[2\pi\left(f_c - \frac{1}{2}f_B\right)t + \pi kt^2 + \frac{3\pi}{4}\right] \quad (6)$$
$$+ J_0(m_1)J_1(m_2)\sin(2\pi f_{LO}t) + DC$$

It should be noted that the high-frequency single-tone microwave signal and high-frequency LFM signal are amplified by an electrical amplifier (EA1) and then sent to the free space through a transmitting antenna. At the same time, the generated IF LFM signal is with a much lower frequency and not in the passband of the transmitting antenna, which can be ignored. Therefore, the transmitted radar signal can be expressed as

$$I_{Trans}(t) = E_1 \cos\left[2\pi\left(f_{LO} + f_c - \frac{1}{2}f_B\right)t + \pi kt^2 + \frac{\pi}{4}\right] + E_2 \sin(2\pi f_{LO} t) \tag{7}$$

where $E_1$ and $E_2$ are amplitudes of the single-chirped LFM signal and the single-tone microwave signal, respectively.

In the receiver, the radar echoes reflected from a moving target are collected by a receiving antenna with a delay time $\Delta\tau$, which can be expressed as

$$\Delta\tau = \frac{2}{c}(R_0 + vt), \tag{8}$$

where $c$ is the velocity of light in a vacuum, $R_0$ is the initial distance, and $v$ is the radial velocity of the target. The sign of $v$ is positive when the target is moving away from the radar system and negative when the target is moving toward the radar system. The echo signal can be expressed as

$$I_R(t) = E_1' \cos\left\{2\pi\left(f_{LO} + f_c - \frac{1}{2}f_B\right)(t - \Delta\tau) + \pi k(t - \Delta\tau)^2 + \frac{\pi}{4}\right\} + E_2' \sin\left[2\pi f_{LO}(t - \Delta\tau)\right] \tag{9}$$

where $E_1'$ and $E_2'$ are the are amplitudes of the single-chirped LFM signal and single-tone microwave signal at the receiving antenna, respectively. The echo signal is amplified by another EA (EA2) and applied to drive a Mach-Zehnder modulator (MZM) via a 90° electrical hybrid coupler to modulate the reference optical signal $E_R(t)$ from the other branch of the PBS. The MZM is biased at the quadrature transmission point to realize SSB modulation. After detecting the optical signal from the MZM in PD2 and filtering it using an electrical low-pass filter (ELPF), two low-frequency signals ($f_{d\text{-}CW}$ and $f_{d\text{-}LFM}$) are generated and shown in Fig. 1(h), which are the de-chirped signal of the LFM signal and the frequency difference of the transmitted and received CW microwave signals. For simplicity, the process of generating the two low-frequency signals is called the de-chirping process. The two low-frequency signals from the ELPF are sampled by a real-time oscilloscope (OSC1) with a low sampling rate and then post-processed by Matlab. The two low-frequency electrical signals can be written as

$$I_{LPF}(t) = A_1 \sin(2\pi f_{LO}\Delta\tau) + A_2 \sin\left[2\pi\left(f_{LO} + f_c - \frac{1}{2}f_B\right)\Delta\tau + \pi kt^2 - \pi k(t - \Delta\tau)^2\right], \tag{10}$$

where $A_1$ and $A_2$ are the amplitudes of the two signals. Therefore, the frequency of the two low-frequency signals can be written as

$$f_{d-CW} = \frac{2|v|}{c} f_{LO}, \tag{11}$$

$$f_{d-LFM} = \frac{2v}{c}\left(f_{LO} + f_C - \frac{1}{2}f_B\right) + kt - k(t - \Delta\tau)\left(1 - \frac{2v}{c}\right). \tag{12}$$

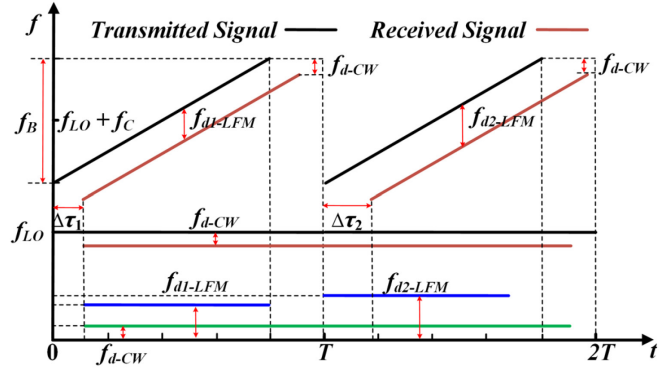

Fig. 2. Schematic diagram of the photonic de-chirping process.

Fig. 2 shows the schematic diagram of the photonic de-chirping process, where $f_{d\text{-}CW}$ is actually the DFS due to the movement of the target, $\Delta\tau_1$ and $\Delta\tau_2$ are the time delays of two adjacent echo signals relative to the transmitted signal, and $f_{d1\text{-}LFM}$ and $f_{d2\text{-}LFM}$ are the corresponding de-chirped frequency. Because the radial velocity of the target is far less than the velocity of light and the value of $(2v/c)\cdot(f_{LO} + f_C + 0.5f_B)$ is far less than $k\Delta\tau$, $f_{d\text{-}LFM}$ in (12) can be simplified as

$$f_{d-LFM} \approx k\Delta\tau . \tag{13}$$

According to (11) and (13), the distance and magnitude of the radial velocity can be derived as

$$R = \frac{1}{2}c\Delta\tau = \frac{c}{2k}f_{d-LFM}, \tag{14}$$

$$|v| = \frac{c}{2f_{LO}}f_{d-CW}. \tag{15}$$

It should be noted that the direction of the radial velocity can be determined according to the distance measured by adjacent reflected echo signals. In practical applications, the frequency change of $f_{d\text{-}LFM}$ caused by the target distance is far beyond the frequency change of $f_{d\text{-}CW}$ caused by the radial velocity. Therefore, assuming that the radial velocity of the target between two adjacent sampled echo signals is constant, the direction of the radial velocity is positive when the measured distance is increasing in the next sampled echo signals, otherwise, the direction of the radial velocity is negative. Therefore, the distance and radial velocity of a moving target can be determined by joint using the two low-frequency de-chirped signals.

In addition to the distance and radial velocity measurements, high-resolution ISAR imaging can also be implemented based on the proposed system. When the ISAR imaging is implemented, a digital high-pass filter (DHPF) is applied to eliminate the influence of the DFS frequency $f_{d\text{-}CW}$ on the high-resolution ISAR imaging. When the radar transmits a sequence of $N$ pulses and each echo signal has $M$ time samples, the de-chirped signal after the DHPF can be rearranged as an $M \times N$ matrix to construct a two-dimension image [8]. Theoretically, the range resolution and the cross-range resolution of the ISAR imaging can be expressed as

$$R_L = \frac{c}{2f_B}, \tag{16}$$

$$R_c = \frac{\lambda}{2T_r\Omega}, \tag{17}$$

where $\lambda$ is the center wavelength of the transmitted LFM signal, $T_r$ and $\Omega$ are the integration time and the rotating speed of the target in a frame image. According to (16) and (17), high-resolution ISAR imaging can be implemented, in which high range resolution is realized by employing a wideband LFM signal, and high cross-range resolution is obtained through a longer integration time or a larger rotational speed of the target.

**3. Experimental results**

*3.1. Experimental setup*

An experiment based on the setup shown in Fig. 1 is carried out to verify the performance of the proposed multi-functional microwave photonic radar system. The CW light wave centered at 1550.698 nm from the LD (HLT-ITLA-M-C-20-1-1-FA) is injected into the DPol-DPMZM (Fujitsu FTM7977) via PC1. Two phase-orthogonal IF LFM signals are generated by an AWG (Keysight M8190A) and sent to the two RF ports of the DPol-DPMZM in the X polarization for SSB modulation. A microwave signal at 8 GHz is generated from a microwave signal generator (MSG, Agilent 83630B) and then sent to a 90° electrical hybrid coupler (Nardar 4065 7.5-16 GHz). The 0° output and the 90° output of the 90° hybrid are sent to the other two RF ports of the DPol-MZM in the Y polarization to realize the CS-SSB modulation of the microwave signal. Following the DPol-DPMZM, the optical signal is split into two branches by a 1:1 optical coupler (OC): one output of the OC is connected to observe the optical spectrum from the DPol-DPMZM, and the other output of the OC is connected to an erbium-doped fiber amplifier (EDFA, Amonics AEDFA-PA-35-B-FA) to amplify the optical signal. The amplified optical signal from the EDFA is sent to a PBS via PC2. By adjusting PC2, the modulated optical signals in the X and Y polarizations of the DPol-DPMZM are superposed and combined as shown in (4) and (5). One output of the PBS is injected into PD1 (DSC-40S, 16 GHz bandwidth) with a responsivity of 0.8 A/W to generate the transmitted signal of the radar system, which contains a high-frequency LFM signal and a high-frequency single-tone microwave signal. The transmitted signal from PD1 is amplified by an EA (CLM 145-7039-293B, 5.85-14.50 GHz), and then radiated through a transmitting antenna (8-18 GHz) for target detection. The echo signals reflected from the targets are collected by a receiving antenna (8-18 GHz), which are amplified by EA2 (CLM 145-7039-293B, 5.85-14.50 GHz). The amplified echo signal is sent to another 90° electrical hybrid coupler (Nardar 4065 7.5-16 GHz) before applied to the MZM (Fujitsu FTM7937) to modulate the reference optical signal from the other output of the PBS. After the SSB-modulated optical signal from the MZM is detected in PD2 (DSC-40S, 16 GHz bandwidth) and filtered by an ELPF (Mini-Circuits SLP-1650, DC-1.4 GHz), the low-frequency de-chirped signal is monitored by OSC1 (Rohde & Schwarz, RTO2032) and then further post processed. In addition, a sampling oscilloscope (OSC2, Agilent DCA-J 86100C) is used to capture the waveform of the generated transmitted signal from PD1. To do so, the AWG is triggered by a 10-MHz reference signal from the MSG. An electrical spectrum analyzer (ESA, Rohde & Schwarz, FSP-40) is used to observe the electrical spectra of the transmitted signal. An optical spectrum analyzer (OSA, ANDO AQ6317B) is used to observe the optical spectra of the key nodes shown in Fig. 1(d), (e), and (f).

*3.2. Optical Spectra from the DPol-DPMZM and the PBS*

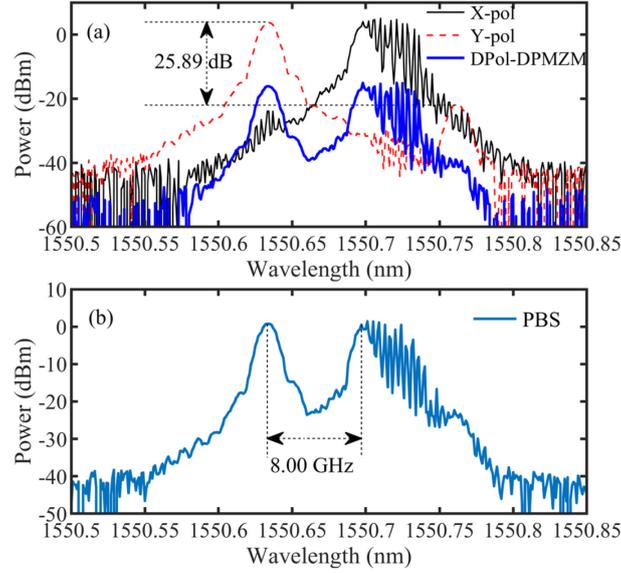

Fig. 3. (a) Optical spectra of the RF modulated optical signals, (b) optical spectrum of the optical signal at one of the PBS outputs.

First, the AWG is used to generate two up-chirped phase-orthogonal IF LFM pulse signals from 0.5 to 4.5 GHz. The width and period of the pulse signals are 80 μs and 100 μs, respectively. The optical spectrum in the X polarization of the DPol-DPMZM after SSB modulation is shown in the black solid line in Fig. 3(a), in which the +1st-order optical sideband of the IF LFM signal and the optical carrier are generated, but cannot be distinguished due to the limited resolution of the OSA. Meanwhile, an 8-GHz single-tone microwave signal is generated from the MSG and only the -1st-order optical sideband of the single-tone microwave signal at 8 GHz is generated in the Y polarization of the DPol-DPMZM, as shown in the red dashed line in Fig. 3(a). The suppression ratio of the -1st-order optical sideband of the single-tone microwave signal is around 25.89 dB. At the output port of the DPol-DPMZM, the two orthogonally polarized optical signals are shown in Fig. 3(a). It should be pointed out here that the modulated optical signals in the X and Y polarizations of the DPol-DPMZM in Fig. 3(a) are measured after the PBS by making the principal axes of the DPol-DPMZM aligned with that of the PBS via tuning PC2.

By adjusting PC2 to make the polarization state of the X-polarized light or the Y-polarized light be oriented to have an angle of 45° to one of the axes of the PBS, the X-polarized light and the Y-polarized light are combined and equally divided by the PBS. The optical spectrum of one PBS output is shown in Fig. 3(b). In Fig. 3(b), the -1st-order optical sideband of the microwave signal is 8 GHz away from the optical carrier, which can be used to determine the exact position of the optical carrier. It should be noted that the two optical signals from the PBS do not have to be completely equally divided to achieve the radar detection functions.

*3.3. Transmitted Radar Signal from PD1*

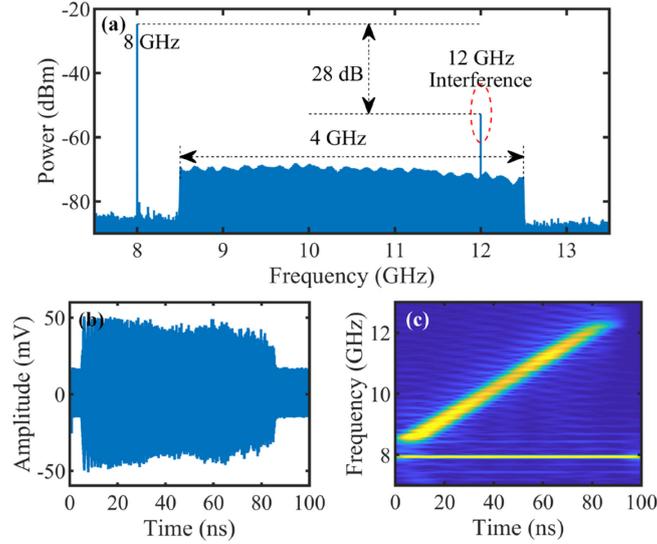

Fig. 4. (a) Electrical spectrum, (b) temporal waveform, and (c) time-frequency diagram of the transmitted radar signal.

The optical signal $E_T(t)$ from one PBS output is sent to PD1 to generate the transmitted radar signal. The electrical spectrum of the transmitted radar signal measured by the ESA with a resolution bandwidth of 10 kHz is given in Fig. 4(a). An LFM signal with a bandwidth of 4 GHz and a single-tone microwave signal at 8 GHz are simultaneously generated. It is seen in Fig. 4(a) that there is an interference signal at 12 GHz which is actually introduced by the AWG due to its nonlinearity. Since the power of the interference signal is much lower than the desired single-tone signal at 8 GHz, its influence can be ignored. In the system, EA1 is used to boost the power of the desired transmitted signal shown in Fig. 4(a), which has a small signal gain of 39 dB and an operating bandwidth from 5.85 to 14.5 GHz. In fact, an LFM signal from 0.5 to 4.5 GHz is also generated at the output of PD1. However, after EA1, the LFM signal from 8.5 to 12.5 GHz and the single-tone microwave signal at 8 GHz are dominant and used as the radar transmitted signal.

To show more detailed information of the radar transmitted signal, OSC2 is used to capture the signal from PD1. When the signal is analyzed, a DHPF is applied to filter out the LFM signal in the low-frequency band. Fig. 4(b) and (c) shows the temporal waveform and time-frequency diagram of the radar transmitting signal in one signal period (100 ns). The time-frequency diagram is obtained by the short-time Fourier transform. It is clearly seen in Fig. 4(c) that an up-chirped LFM signal and a single-tone microwave signal are generated. It should be noted that the pulse period of the signal in this measurement is decreased from 100 μs to 100 ns with other parameters unchanged because of the limited sampling points of OSC2. In the following parts for distance and velocity measurement and high-resolution microwave imaging, the period of the signal is 100 μs.

### 3.4. Measurement of Distance and Radial Velocity

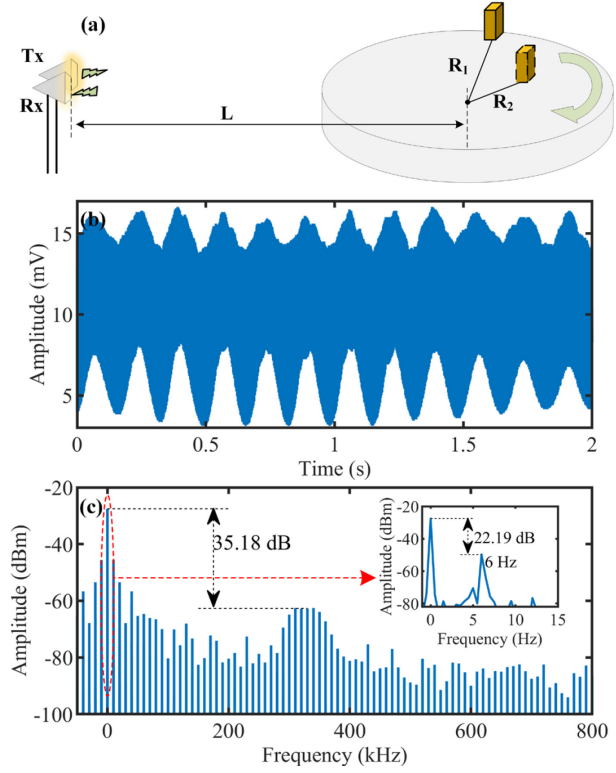

Fig. 5. (a) The schematic diagram of the target detection, (b) temporal waveform and (c) electrical spectrum of the de-chirped signal at the output of the ELPF. Tx: transmitter; Rx: receiver.

In the receiver, the radar echoes reflected from a rotating target are amplified by EA2 and then sent to the MZM to modulate the reference optical signal $E_R(t)$. The optical signal from the MZM is detected in PD2 to generate the desired de-chirped signal. The de-chirped signal is captured using OSC1 and then post-processed using Matlab. In the measurement of distance and radial velocity, a cuboid is used as the rotating target. The length, width, and height of the target are 10, 8, and 18 cm, respectively. The target is located on a turntable and is $R_1$=45 cm away from the center of the turntable, as shown in Fig. 5(a). Along the radar line of sight, the center of the turntable is about $L$=1.32 m away from the radar antenna pair. In the experiment, the turntable is rotating in the horizontal plane and its period is about 24.56 s. According to (13), the maximum frequency of the de-chirped signal is less than 600 kHz. Therefore, the de-chirped signal is sampled by OSC1 with a sampling rate of 4 MHz. During the rotation, the echoes are sampled every 17/16 period with 2 s sampling time for a total of 17 samples realized by using OSC1 which is controlled by a computer. The cuboid is at the same position at the first and the last samples. Fig. 5(b) shows the temporal waveform of the fourth sample. It can be seen that the de-chirped signal has a sampling time of 2 s and the corresponding frequency resolution in the post-processing is 0.5 Hz. Fig. 5(c) shows the spectrum obtained by performing the fast Fourier transform (FFT) to the sampled signal. According to (11) and (13), the DFS caused by the rotation of the target is less than 8 Hz, while the maximum frequency of the de-chirped signal is less than 600 kHz taking the time delay between transmitted and echo signals into account after range calibration, which is consistent with the experimental result in Fig. 5(c). Therefore, the distance and magnitude of the radial velocity can be simultaneously obtained. It should be noted that the influence of the DC component from PD2 has been eliminated in the measurement of velocity, which increases the measurement error of the velocity when the radial velocity of the target is close to zero.

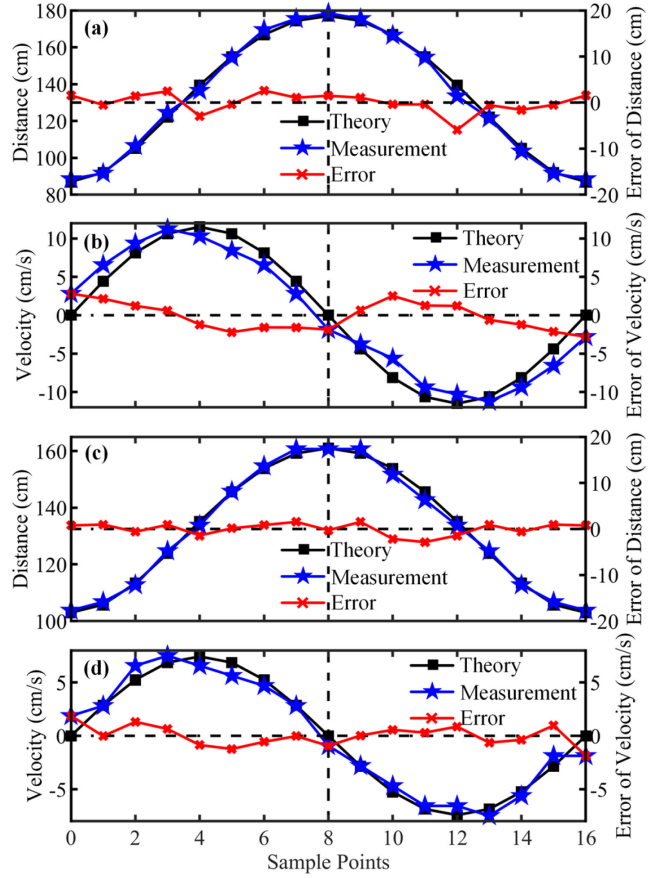

Fig. 6. (a) Distance and (b) radial velocity measurement results when the target is $R_1$=45 cm away from the center of the turntable, (c) distance and (d) radial velocity measurement results when the target is $R_2$=29 cm away from the center of the turntable.

To determine the direction of the radial velocity at a specific time, not only the sampled echo signals at the specific time but also the following sampled echo signals need to be used. The direction of the radial velocity is positive when the measured distance is increasing in the next sampled signals, otherwise, the direction of the radial velocity is negative. According to (14) and (15), the distances and radial velocities of the target are extracted from the de-chirped low-frequency signals.

Two experiments are carried out to show the capability of distance and radial velocity measurement. In the first scenario, the cuboid is $R_1$=45 cm away from the center of the turntable. Distance and radical velocity measurement results are shown in Fig. 6(a) and (b), respectively. The blue stars represent the measured value of the distances and radial velocities, the black squares denote the theoretical value, and the red crosses are the measurement errors. In another scenario, the target is $R_2$=29 cm away from the center of the turntable with other parameters unchanged. Distance and radical velocity measurement results are shown in Fig. 6(c) and (d), respectively. From Fig. 6 we can see that the measured values of the distances and radial velocities and the theoretical values agree well with each other. The absolute errors of distances and radial velocities are also measured, which are small than 5.9 cm and 2.8 cm/s, respectively.

*3.5. High-resolution ISAR Imaging*

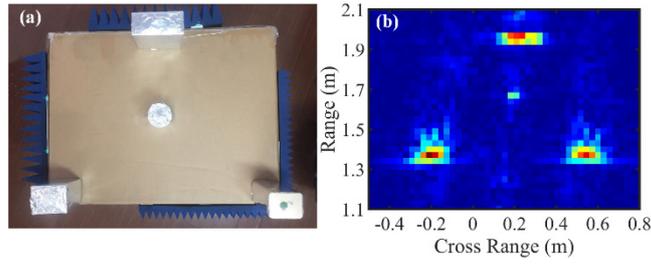

Fig. 7. (a) The photograph of four targets under test, (b) the imaging result of the four targets.

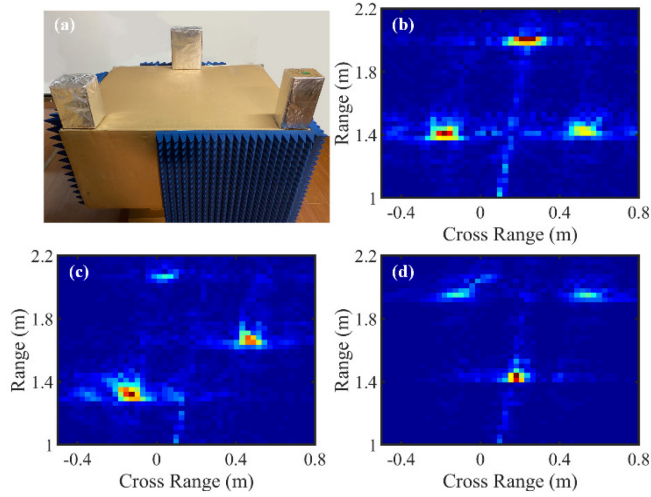

Fig. 8. (a) The photograph of three cuboids packed with silver papers, (b), (c), and (d) are the imaging results for the first, fourth, and seventh samples, respectively.

With the proposed multi-functional microwave photonic radar system, not only can the distance and radial velocity of a rotating target be measured, but also high-resolution ISAR imaging can be implemented. In the first experiment, the transmitted radar signals are a positive chirp LFM signal with a bandwidth of 4 GHz (8.5 to 12.5 GHz) and a single-tone microwave signal at 8 GHz. Three cuboids and a cylinder as the targets are placed in different positions on the turntable, which is shown in Fig. 7(a). The turntable is rotating in the horizontal plane with a period of 24.56 s. Along the radar line of sight, the center of the turntable is placed at a distance of about 1.67 m away from the antenna pair. According to (13), the de-chirped signal is sampled with a sampling rate of 4 MSa/s and a sampling time of 2 s. A DHPF is applied to eliminate the influence of the DFS frequency $f_{d\text{-}CW}$ on the ISAR imaging. According to (16) and (17), the theoretical range resolution and the cross-range resolution are 3.75 cm and 2.79 cm, respectively. Fig. 7(b) shows the imaging result of the four targets. As can be seen, three cuboids and one smaller cylinder can be clearly distinguished. Therefore, high-resolution ISAR imaging can be achieved with a sampling rate of 4 MSa/s.

To further verify the performance of the multi-functional radar system with real-time imaging ability, three cuboids packed with silver paper are used as the targets, which are placed on the turntable shown in Fig. 8(a). Difference from the parameters of the first experiment, the distance between the center of the turntable and the radar antenna pair is about 1.73 m. During the rotation, the echoes are sampled every 13/12 period with 2 s sampling time for a total of 13 samples realized by using OSC1 which is controlled by a computer. The de-chirped signal is sampled at 4 MSa/s. Fig. 8(b) shows the imaging result of the first sample. Three targets can be also clearly distinguished. Fig. 8(c) and (d) shows the imaging results corresponding to the fourth sample and the seventh sample, respectively. As can be seen, the real-time imaging capability of the multi-functional radar system is verified.

## 4. Discussions

### 4.1. Tunability of the System

The ability to have diverse working modes of the multi-functional radar system is of vital importance in military affairs and civilian activities. To verify the tunability of the proposed system, microwave signals of different frequencies and IF LFM signals with different bandwidths are used to generate various forms of the transmitted radar signal.

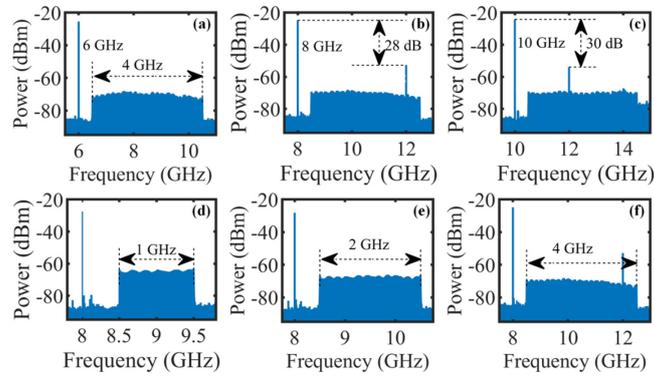

Fig. 9. Electrical spectra of the generated radar signal with frequency and bandwidth tunability.

First, the IF LFM signal is fixed from 0.5 to 4.5 GHz and the single-tone microwave signal varies from 6 to 10 GHz. After photonic frequency up-conversion, the transmitted radar signals with the same LFM bandwidth and different frequencies are generated, which are shown in Fig. 9(a), (b), and (c). It should be pointed out here that there is an interference signal at 12 GHz which is actually introduced by the AWG due to its nonlinearity. To further verify that the radar system has diverse working modes, the single-tone microwave signal at 8 GHz is fixed, while the bandwidth of the IF LFM signal varies from 1 to 4 GHz. After photonic frequency up-conversion, the transmitted radar signals with different bandwidths are generated and shown in Fig. 9(d), (e), and (f).

### 4.2. Resolution of the ISAR Imaging

In modern military affairs and civilian activities, it is important to acquire the shape information of the targets. In the experiment, a 4-GHz bandwidth LFM signal is applied and the theoretical range resolution is as high as 3.75 cm. The range resolution can be further improved by using a radar signal with a larger bandwidth. As for the cross-range resolution, it is determined by the center frequency of the LFM radar signal, the integration time, and the rotational speed of the target. In the experiment, the integration time of the targets is as long as 2 s to increase the cross-range resolution. However, the integration time cannot be increased arbitrarily, because a longer integration time will reduce the real-time imaging performance of the radar system. A larger rotational speed may provide higher cross-range resolution when the integration time of the targets and the center frequency of the radar signal keep constant. As a matter of fact, the rotational speed of the target is 0.256 rad/s, which is limited by our turntable and leads to a cross-range resolution of 2.79 cm. In fact, the cross-across resolution and the real-time imaging performance can be further improved by increasing the rotational speed of the targets.

*4.3. Comparison of Different Methods*

The key feature and differences between the proposed system and those reported before are further discussed. To the best of our knowledge, the proposed distance and radial velocity measurement method using only a single-chirped LFM signal and a single-tone microwave signal has never been reported before in microwave photonic radar systems. Some methods based on photonic approaches have been reported in the last few years to measure the distance and radial velocity of a moving target [22]-[24], which generally use LFM signals with opposite chirps containing up-chirped and down-chirped segments. In [22], the distance of a single stationary target is measured, while the distance and radial velocity measurement of the moving target has not been obtained in the experiment. In [23] and [24], the distance and radial velocity information of a moving target is obtained by utilizing the LFM signals with opposite chirps. However, in their experiments, the echo signal is not reflected by the moving target but is directly generated by the AWG. In practical applications, the DFS caused by the radial velocity is usually small, which makes the two de-chirped frequencies generated by the up-chirped and down-chirped LFM signals be difficult to distinguish in the frequency domain, which may reduce the distance and radial velocity measurement accuracy. Furthermore, the very close two de-chirped frequencies make the system difficult to image. In comparison, the distance and radial velocity measurement approach proposed in this paper is based on a single-chirped LFM signal and a single-tone microwave signal. The single-chirped LFM signal is used to measure the distance of a target, while the single-tone microwave signal is used to measure the magnitude of its radial velocity. The direction of the radial velocity can also be determined by simply analyzing the adjacent sampled echo signals. Generally, the low-frequency signal corresponding to the DFS and the de-chirped frequency from the LFM signal are separated in the frequency domain. By filtering out the low-frequency signal corresponding to the DFS, high-resolution ISAR imaging can be implemented by using the de-chirped frequency from the LFM signal.

## 5. Conclusions

In summary, we have proposed and experimentally demonstrated a multi-functional microwave photonic radar system for simultaneous distance and radial velocity measurement and high-resolution microwave imaging. The key significance of the work is that, for the first time, a composite transmitted microwave signal of a single-chirped LFM signal and a single-tone microwave signal is used in a microwave photonic radar system to realize both distance and velocity measurement and high-resolution microwave imaging, which solves the problems that the conventional microwave photonic single-chirped LFM radar system cannot measure the velocity and the microwave photonic dual-chirped LFM radar system is difficult to image. An experiment is performed. An up-chirped LFM signal from 8.5 to 12.5 GHz and an 8.0 GHz single-tone microwave signal are generated and used as the transmitted radar signal. The absolute measurement errors of distance and radial velocity are less than 5.9 cm and 2.8 cm/s, respectively. ISAR imaging results are also demonstrated, which proves the high-resolution and real-time ISAR imaging ability of the proposed system. Because of its simple structure and different functions that can be implemented simultaneously, the proposed system can provide a new photonics-based solution for multi-function radar systems.

**Funding**

Natural Science Foundation of Shanghai (20ZR1416100); National Natural Science Foundation of China (NSFC) (61971193); Open Fund of State Key Laboratory of Advanced Optical Communication Systems and Networks, Peking University, China (2020GZKF005), Science and Technology Commission of Shanghai Municipality (18DZ2270800).

**Conflicts of interest**

The authors declare no conflicts of interest.